\documentstyle[prd,aps,epsfig,floats]{revtex}
\begin{document}
\draft
\wideabs{
\title{Metropolis Importance Sampling for Rugged Dynamical Variables}

\author{ Bernd A. Berg$^{\rm \,a,b}$ }

\address{ (E-mail: berg@csit.fsu.edu)\\ 
$^{\rm \,a)}$ Department of Physics, Florida State University,
  Tallahassee, FL 32306\\
$^{\rm \,b)}$ School of Computational Science and Information 
Technology, Florida State University, Tallahassee, FL 32306\\
} 

\date{September 16, 2002; revised March 20, 2003}

\maketitle
\begin{abstract}

A funnel transformation is introduced, which acts recursively from 
higher towards lower temperatures. It biases the a-priori probabilities 
of a canonical or generalized ensemble Metropolis simulation, so that 
they zoom in on the global energy minimum, if a funnel exists indeed. 
A first, crude approximation to the full transformation, called rugged 
Metropolis one (RM$_1$), is tested for Met-Enkephalin. At 300$\,$K the 
computational gain is a factor of two and, due to its simplicity, 
RM$_1$ is well suited to replace the conventional Metropolis updating 
for these kind of systems. 
\end{abstract}
}

\narrowtext

To explain important aspects of protein folding, Bryngelson and Wolynes 
introduced a funnel picture\cite{BrWo87}, which is supported by various 
numerical results~\cite{SoOnWo96,HOO99,HaOn01}. Nevertheless, the 
understanding as well as the practical relevance of the funnel concept 
has remained somewhat limited. The reason is that the funnel lives in 
the high-dimensional configuration space, while numerical studies have 
been confined to projections onto so called reaction coordinates, and 
there is no generic definition of a good reaction coordinate. Here I 
follow a different path and introduce a general funnel description 
from higher towards lower temperatures. It yields a powerful new 
method for designing Metropolis~\cite{Me53} weights.

In protein models the energy $E$ is a function of a number of 
dynamical variables $v_i,\, i=1,\dots,n$, whose fluctuations in the 
Gibbs canonical ensemble are described by a probability density~(pd) 
$\rho ( v_1,\dots ,v_n; T)$, where $T$ is the temperature.
To be definite, we use in the following the all-atom energy
function~\cite{SNS84} ECEPP/2 (Empirical Conformational Energy 
Program for Peptides).
Our dynamical variables $v_i$ are the dihedral angles, each chosen 
to be in the range $-\pi\le v_i< \pi$, and the volume of the 
configuration space is $K=(2\pi)^n$.

Let us define the {\it support} of a pd of the dihedral angles.
Loosely speaking, the support of a pd is the region of configuration 
space where the protein wants to be. Mathematically, we define $K^p$ 
to be the smallest sub-volume of the configuration space for which
\begin{equation} \label{support}
 p = \int_{K^p}\, \prod_{i=1}^n d\,v_i\, \rho (v_1,\dots,v_n;T)
\end{equation}
holds. Here $0<p<1$ is a probability, which ought to be chosen close 
to one, e.g., $p=0.95$. The free energy landscape at temperature 
$T$ is called {\it rugged}, if the support of the pd consists of many 
disconnected parts (this depends of course a bit on the adapted values 
for $p$ and ``many''). 
That a protein folds at room temperature, say $300\,K$, into a
unique native structure $v^0_1,\dots,v^0_n$ means that its pd
$\rho ( v_1,\dots ,v_n; 300\,K)$ describes small fluctuation
around this structure. 
We are now ready to formulate the funnel picture in terms of pds.
Let us choose a protein and consider for it a sequence of pds 
\begin{equation} \label{rho_r}
\rho_r ( v_1,\dots ,v_n) = \rho ( v_1,\dots ,v_n; T_r),\ 
r=1,\dots,s\ , 
\end{equation}
which is ordered by the temperatures $T_r$, namely
\begin{equation} \label{T_order}
T_1\ >\ T_2\ >\ \dots\ >\ T_f\ .
\end{equation}
The sequence~(\ref{rho_r}) constitutes a protein {\it funnel} when,
for a reasonable choice of the probability $p$ and the 
temperatures~(\ref{T_order}), the following holds:

\begin{enumerate}

\item The pds are rugged.

\item The support of a pd at lower temperature is contained in the 
      support of a pd at higher temperature
\begin{equation} \label{Kp_order}
K^p_1\ \supset\ K^p_2\ \supset\ \dots\ \supset\ K^p_f\ ,
\end{equation}
e.g. for $p=0.95$, $T_1=400\,K$ and $T_f=300\,K$. 

\item With decreasing temperatures $T_r$ the support $K^p_r$ 
      shrinks towards small fluctuations around the native structure.

\end{enumerate}

Properties 2 and 3 are fulfilled for many systems of statistical 
physics, when some groundstate stands in for the native structure.
The remarkable point is that they may still hold for certain complex 
systems with a rugged free energy landscape, i.e., with property~1 
added. In such systems one finds typically local free energy minima,
which are of negligible statistical importance at low temperatures, 
while populated at higher temperatures. In simulations at low 
temperature the problem of the molecular dynamics (for a review 
see~\cite{Frenkel}) as well as of the Metropolis~\cite{Me53} 
canonical ensemble approach is that the updating tends to get stuck 
in those local minima. On realistic simulation time scales this 
prevents convergence towards the native structure. 
On the other hand, the simulations move quite freely at higher 
temperatures, where the native structure is of negligible 
statistical weight. Nevertheless, the support of a protein pd 
may already be severely restricted, as we shall illustrate.
The idea is to use a relatively easily calculable pd at a higher 
temperature to improve the performance of the simulation at a lower 
temperature. In the following we investigate this idea for the
Metropolis algorithm. 

The Metropolis importance sampling would be perfected, if we could 
propose new configurations $\{v_i'\}$ with their canonical pd 
$\rho (v'_1,\dots,v'_n;T)$. 
Due to the funnel property~2 we expect that an {\it estimate}
${\overline \rho} (v_1,\dots,v_n;T')$ from some sufficiently close-by
higher temperature $T'>T$ will feed useful information into the 
simulation at temperature $T$. The potential for computational gains 
is large because of the funnel property~3. The suggested scheme for 
the Metropolis updating at temperature $T_r$ is to propose new 
configurations $\{v_i'\}$ with the pd~(\ref{rho_r}) 
${\overline \rho}_{r-1} (v'_1,\dots,v'_n)$ and to accept them 
with the probability
\begin{equation} \label{P0_acpt}
P_a = \min \left[ 1, \exp \left( - {E'-E\over k\,T_r} \right)\,
{ {\overline \rho}_{r-1} (v_1,\dots,v_n)\over 
{\overline \rho}_{r-1} (v'_1,\dots,v'_n)} \right]\ .
\end{equation}
This equation biases the a-priori probability of each dihedral 
angle with an estimate of its pd from a higher temperature. In
previous literature~\cite{Br85,MHB86} such a biased updating
has been used for the $\phi^4$ theory, where it is efficient to
propose $\phi(i)$ at each lattice size $i$ with its single-site 
probability.

For our temperatures $T_r$ the ordering~(\ref{T_order}) is 
assumed. With the definition 
${\overline \rho}_0 (v_1,\dots,v_n) = (2\pi)^{-n}$
the simulation at the highest temperature, $T_1$, is performed 
with the usual Metropolis algorithm. We have thus a recursive 
scheme, called rugged Metropolis (RM) in the following. When
${\overline \rho}_{r-1} (v_1,\dots,v_n)$ is always a usefull 
approximation of $\rho_r (v_1,\dots,v_n)$, the scheme 
zooms in on the native structure, because the pd at $T_f$ governs 
its fluctuations. 

To get things working, we need to construct an estimator 
${\overline \rho} (v_1,\dots,v_n;T_r)$ from the numerical data 
of the RM simulation at temperature $T_r$. Although this is 
neither simple nor straightforward, a variety of approaches offer 
themselves to define and refine the desired estimators. In the 
following we work with the approximation
\begin{equation} \label{rho0_T0}
{\overline \rho} (v_1,\dots,v_n;T_r) = \prod_{i=1}^n
{\overline \rho}^1_i (v_1,\dots,v_n;T_r) 
\end{equation}
where the ${\overline \rho}^1_i (v_1,\dots,v_n;T_r)$ are estimators 
of reduced one-variable pds defined by
\begin{equation} \label{pd1}
 \rho^1_i ( v_i; T) = \int_{-\pi}^{+\pi} \prod_{j\ne i} d\,v_j\,
 \rho ( v_1,\dots ,v_n; T)\ .
\end{equation}
The resulting algorithm, called RM$_1$, appears to constitute the 
simplest RM scheme possible. Its implementation is straightforward, 
as estimators of the one-variable reduced pds are easily obtained 
from the time series of a simulation. The computer time consumption 
of RM$_1$ is practically identical with the one of the conventional 
Metropolis algorithm. In the following RM$_1$ is used to 
demonstrate the correctness of our basic assumptions.

The scope of this paper limits us to one illustration. We rely on
a simulation of the brain peptide Met-Enkephalin, because it is a 
numerically well studied system, which allows for comparison with 
results of the literature~\cite{LiSh87,OkKi92,HaOk93,MeMe94}. Our
Metropolis simulations are performed with a variant of SMMP~\cite{Ei01}
(Simple Molecular Mechanics for Proteins). We keep the $\omega$
torsion angles unconstrained and thus have 24 fully variable 
dihedral angles. In the previous literature the omega angles were 
either fixed to $\pi$ or restricted to $[-\pi + \pi/9,\pi-\pi/9]$. 
This leads to statistically significant differences in the 
energy, while the structural differences remain negligible.
The performance of the RM$_1$ updating was tested
at 300$\,$K using input from a simulation at 400$\,$K. The value of
300$\,$K is chosen, because it is in the temperature range on 
which simulations of biological molecules eventually have to focus.
The temperature of 400$\,$K is high enough so that the
Metropolis algorithm is efficient as the autocorrelation times are 
small, while it is low enough to provide useful input for the 
300$\,$K simulation.

At each simulation 
temperature a time series of $2^{17}=131,072$ configurations is 
kept, in which subsequent configurations are separated by 32 sweeps. 
A sweep is defined by updating each dihedral angle once.
Before starting with the measurements 
$2^{18}=262,144$ sweeps are performed for reaching equilibrium. Thus, 
the entire simulation at one temperature relies on 
$2^{18}+2^{22}=4,456,448$ sweeps. On a modern PC 
($1.9\,$GHz Athlon) this takes under 12 hours for the vacuum system and 
less than two days with the inclusion~\cite{OONS87} of solvent effects.
For each dihedral angle the acceptance rate of the Metropolis 
algorithm was monitored at run time and the integrated autocorrelation 
time $\tau_{\rm int}$ (see~\cite{Sokal} for its definition) was 
calculated from the recorded time series. 
Values around 0.5 are desirable for the acceptance rate, but
the decisive quantity for the performance of an algorithm 
is the integrated autocorrelation time. To achieve a pre-defined 
accuracy, the computer time needed is directly proportional to 
$\tau_{\rm int}$. 
In the following results from vacuum simulations are summarized.
Computations which include solvent effects will be reported 
elsewhere~\cite{BHG}. 

\begin{table}[ht]
\caption{ Acceptance rates and integrated autocorrelations times for 
the energy $E$ and, in the SMMP notation, the dihedral angles 
Gly-2~$\omega$ ($v_6$) and Gly-3~$\phi$ ($v_{10}$). 
\label{tab_Met_dia} }
\medskip
\centering
\begin{tabular}{||c|c|c|c|c|c||}   
 Angle  &Method    &400$\,K$&400$\,K$ &300$\,K$&300$\,$K   \\ \hline
        &          & acpt   &$\tau_{\rm int}$
                                      & acpt   &$\tau_{\rm int}$
                                                              \\ \hline
$E$     &Metro     & 0.168  &4.98 (20)& 0.120  & 49.6 (5.0)\\ \hline
$E$     &RM$_1$    & $-$    & $-$     & 0.375  & 26.2 (1.6)\\ \hline
$E$     &PT        & 0.167  &3.67 (20)& 0.119  & 19.9 (1.6)\\ \hline
$E$     &PT+RMC$_1$& 0.460  &2.56 (34)& 0.375  & 9.94 (60) \\ \hline
$v_6 $  &Metropolis& 0.049  &3.09 (10)& 0.034  &21.1 (1.8) \\ \hline
$v_6 $  &RM$_1$    & $-$    & $-$     & 0.416  & 9.68 (66) \\ \hline
$v_6 $  &PT        & 0.049  &2.24 (07)& 0.034  & 7.85 (36) \\ \hline
$v_6 $  &PT+RM$_1$ & 0.553  &1.34 (04)& 0.413  & 4.62 (55) \\ \hline	
$v_{10}$&Metro     & 0.088  &7.49 (47)& 0.034  & 167 (27)  \\ \hline
$v_{10}$&RM$_1$    & $-$    & $-$     & 0.070  & 80.6 (7.0)\\ \hline
$v_{10}$&PT        & 0.087  &6.13 (30)& 0.034  & 32.7 (3.1)\\ \hline
$v_{10}$&PT+RMC$_1$& 0.141  &4.43 (26)& 0.070  & 22.6 (2.7)\\
\end{tabular} \end{table} 

In table~\ref{tab_Met_dia} results for the energy and two dihedral 
angles are presented. Error bars are given in parenthesis. The 
acceptance rates are accurate to $\pm 1$ in their last digit. 
Using the SMMP~\cite{Ei01} conventions, which differ
from previous literature~\cite{LiSh87,OkKi92,HaOk93}, the angles 
are Gly-2~$\omega$ ($v_6$) and Gly-3~$\phi$ ($v_{10}$). They are 
well-suited to illustrate important features of our approach. 

\begin{figure}[-t] \begin{center}
\epsfig{figure=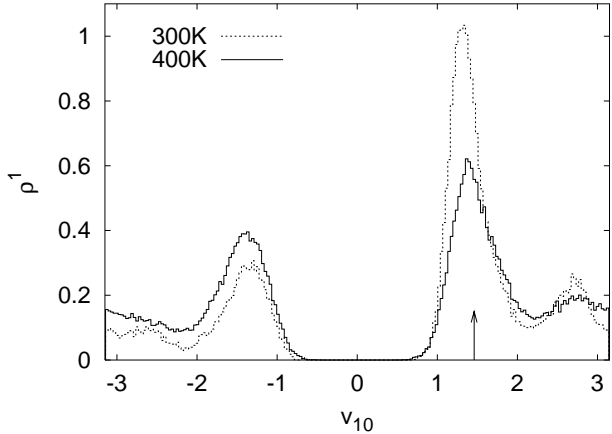,width=\columnwidth} \vspace{-1mm}
\caption{Vacuum probability densities at $400\,K$ and $300\,K$ 
for the Met-Enkephalin dihedral angle Gly-3 $\phi$.
\label{fig_Met_dia10} }
\end{center} \vspace{-3mm} \end{figure}

The Gly-3~$\phi$ angle and four more angles (Gly-2~$\phi$, 
Gly-2~$\psi$, Phe-4~$\phi$ and Phe-4~$\psi$) exhibit very large 
autocorrelation times.
In figure~\ref{fig_Met_dia10} estimates of the one-variable 
pds~(\ref{rho0_T0}) for the Gly-3 $\phi$ angle at 400$\,K$ and 
300$\,K$ are depicted.  The rugged nature of the distribution 
is already obvious at 400$\,K$. While the shapes of the other 
23 dihedral angle pds vary greatly, featuring 
from one to three local maxima, a ruggedness and our funnel 
property~2 are found for each case. Towards low temperatures a 
shrinking of the one-variable pds to their global energy minimum
(GEM) implies property~3. The arrow in 
figure~\ref{fig_Met_dia10} indicates the GEM value of this 
particular angle. Note that for the Metropolis algorithm 
the $\rho^1_i\approx 0$ regions 
do {\it not} constitute free energy barriers, because they 
can be jumped by a single Metropolis updating step. 
This is different for molecular dynamics simulations.

Despite the similarities of the Gly-3 $\phi$ pds at 400$\,$K and 
300$\,$K, there is a big increase of the Gly-3~$\phi$ integrated 
autocorrelation time.  In the conventional, canonical Metropolis 
simulation it is by more than a factor of twenty, from 7.5 to 167.
The RM$_1$ updating reduces this by a factor of 2.1, from 167 to 
81. The integrated autocorrelations times of 
table~\ref{tab_Met_dia} are given in units of 32 sweeps, as this
is the step-size of our time series data recorded.

The acceptance rates for moves of the Gly-3~$\phi$ angle are very
small. As the support of its pds in figure~\ref{fig_Met_dia10} 
covers more than 50\% of the full range, this has to be due to 
correlations with other dihedral angles. To exhibit a rather
distinct case of an angle with a low acceptance rate, results
for one of the six $\omega$ angles are also included in 
table~\ref{tab_Met_dia}. For the conventional Metropolis 
simulation this angle has the same acceptance rate as the 
Gly-3~$\phi$ angle (incidentally to all digits given). A look 
at the Gly-2~$\omega$ pds reveals an obvious reason: They are 
narrowly peaked around the (identified) values $\pm \pi$, which
is explained by the specific electronic hybridization
of the CO-N peptide bond.
Autocorrelations times are about eight times smaller 
than for the Gly-3~$\phi$ angle. Applying RM$_1$ updating to 
the Gly-2~$\omega$ pds cures entirely the problem of 
its low acceptance rate. At 300$\,$K the increase is from 0.034 
to 0.416 and similar numbers are found for the other $\omega$ 
angles. In contrast to that the Gly-3~$\phi$ acceptance rate 
increases only to the modest value of 0.07, while the 
improvements of $\tau_{\rm int}$ are kind of similar. For 
Gly-2~$\omega$ it is by a factor of 2.7 from 21 to 7.9. 

It is straightforward to combine the RM$_1$ updating with 
generalized ensemble methods (see~\cite{MSO01} for a review).
They are enabling techniques for studying equilibrium physics 
of complex systems at very low temperatures~\cite{BeCe92}. 
In the parallel tempering (PT) approach~\cite{Ge91,HuNe96} 
two or more replica are simulated in parallel at distinct 
temperatures and Metropolis exchanges of the temperatures 
are offered occasionally. Autocorrelations are reduced, 
when suitable excursions to higher temperatures become feasible.  
In table~\ref{tab_Met_dia} results for a PT simulation with 
replica at 400$\,$K and 300$\,$K are included. To compare with 
RM$_1$, we use the integrated autocorrelation time of the 
energy, which characterizes the over-all performance.
The PT method decreases $\tau_{\rm int}$ by a divisor of 
2.5, from 50 to 20. When parallel nodes with a reasonably fast
communication are available, this is the gain in real time.
For RM$_1$ the improvement factor is 1.9, from 50 to 26.
Thus, PT outperforms RM$_1$ in real time, while RM$_1$ wins 
in the total CPU time. Most remarkable is that the 
improvement factors multiply. Running PT with the RM$_1$ 
updating yields another factor of two, $\tau_{\rm int}$ is 
reduced from 20 to 10 and, altogether, from 50 to 10.
Note that the acceptance rates are not changed by PT.

\begin{figure}[-t] \begin{center}
\epsfig{figure=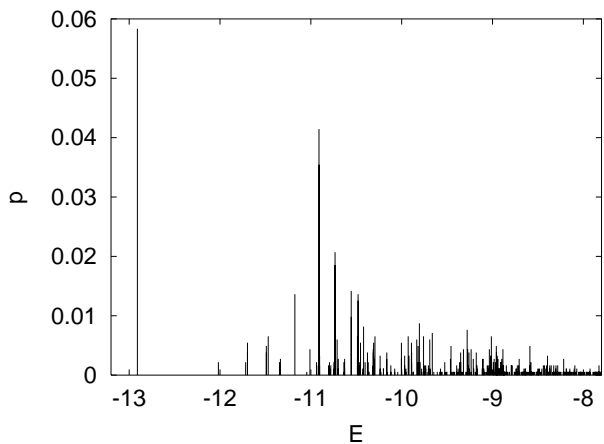,width=\columnwidth} \vspace{-1mm}
\caption{Spectrum of energy-minimized Met-Enkephalin configurations
from the $300\,K$ RM$_1$ simulation in vacuum. \label{fig_Met_emin} }
\end{center} \vspace{-3mm} \end{figure}

In previous simulations~\cite{HaOk93,HOO99} it has been shown 
that considerably lower temperatures than 300$\,$K are needed 
to reduce the fluctuation of Met-Enkephalin in vacuum to 
fluctuation around its GEM. Here, I like to point out that 
each of our simulations at 300$\,$K reaches the valley of 
attraction of the GEM sufficiently often, so that the GEM 
can be found by local minimization. To be specific, the 
energy spectrum of figure~\ref{fig_Met_emin} is obtained from our 
RM$_1$ simulation at $300\,K$ in the following way: First 
we isolate all configurations of the time series which are minima 
in the lower 10\% q-tile of the energy distribution and separated 
by an excursion of the time series into the upper 10\% q-tile.
In this way we obtain 1357 configurations and the SMMP minimizer 
is run on each of them. The probabilities of the resulting spectrum 
levels are plotted in figure~\ref{fig_Met_emin}. The GEM occurs 
at $E=-12.91\,$Kcal/mol with about 6\% probability (107 times) 
followed by a conformation at $E=-10.92\,$Kcal/mol with about 
4\% probability. The frequency of finding the GEM for the other 
simulations of table~\ref{tab_Met_dia} is approximately 
proportional to $\tau_{\rm int}^{-1}$ of the energy. 

Rugged distributions of the dynamical variables are typical for 
Metropolis simulations of proteins and RM$_1$ improves the 
importance sampling. It ought to become standard, as it yields
a relevant gain in computer time for little extra efforts by
the programmer.
Met-Enkephalin is essentially solved by a simulation at $300\,K$. 
For some of its dihedral angles the RM$_1$ updating overcomes the 
problem of low acceptance rates entirely. For others the improvement 
remains more modest, because their low acceptance rates are due to 
correlations with other angles. Only multi-variable moves can 
achieve importance sampling in such a situation. 
For the usual Metropolis algorithm the acceptance 
rates for multi-variable moves are practically zero. Here, we 
gained novel physical insight into the funnel picture, which 
provides us with a recipe on how to design multi-variable moves,
so that the acceptance rate is expected to stay reasonably large. 
One will start with the angles with the worst autocorrelations 
and construct their two-angle moves according to their reduced 
two-variable pds, the RM$_2$ algorithm. Next, one may use
three variables, and so on. In this 
way the outcome of our investigation of computational consequences 
of the funnel picture is that we do not expect a single, generically 
good Metropolis algorithm for protein simulations. Instead, we have 
developed a strategy to design an algorithm for each particular 
protein. This aspect of the RM approach promises advances 
for simulations of larger peptides.

\acknowledgments
I would like to thank Yuko Okamoto for numerous useful discussions.
This research was in part supported by the Okazaki National Institutes 
and by the U.S. Department of Energy under contract DE-FG02-97ER41022.

\clearpage
\end{document}